# Galaxies with Declining Rotation Curves

D. I. Zobnina[a], * and A. V. Zasov[b, c], **

[a] *Astro Space Center, Lebedev Physical Institute, Russian Academy of Sciences, Moscow, Russia*
[b] *Sternberg Astronomical Institute, Moscow State University, Moscow, Russia*
[c] *Faculty of Physics, Moscow State University, Moscow, Russia*
*e-mail: zobninadaria@mail.ru
**e-mail: a.v.zasov@gmail.com

**Abstract**—A sample of 22 spiral galaxies compiled from published data is studied. The galaxy rotation curves pass through a maximum distance of more than ~1 kpc from the center with a subsequent decrease in the rotation velocity. The galaxy position in the Tully–Fisher (TF) and baryonic Tully–Fisher (BTF) diagrams show that the values of maximum rotation velocities are located on the same sequence with other galaxies, while the velocities at the disk periphery for some galaxies are significantly lower than the expected values for a given mass or luminosity. Thus, the decrease in the rotation curve can be associated with a reduced contribution of the dark halo to the rotation velocity. For seven galaxies with the longest rotation curves, the disk mass was estimated to be with the dark halo (Newtonian model) and without the halo (modified Newtonian dynamics (MOND) model). In four of the galaxies, the MOND model encounters difficulties in interpreting the rotation curve: in order to be consistent with the observations, the MOND parameter $a_0$ should differ significantly from the expected value $a_0 \sim 10^{-8}$ cm/s2, while the disk mass exceeds the value based on IR photometry and maximum disk model. The conflict with MOND is the most significant for NGC 157.

## 1. INTRODUCTION

Radial dependences of the galactic disk's circular rotation velocity (rotation curves) $V(R)$ with increasing $R$ generally reach an extended flat segment characterized by a nearly constant rotation velocity $V_{\text{far}}$. However, in some cases, the rotation velocity increases or decreases at large distances from the center (see, e.g., [1–4]).

The rotation curve shape is known to correlate with the optical characteristics of galaxies (see the discussion of this topic in [5]). Galaxies with declining rotation curves, as a rule, have high luminosity and surface brightness, while continuously increasing $V(R)$ is usually found in dwarf galaxies (see, e.g., [6], where the authors present a "universal rotation curve" that depends only on the galaxy's total luminosity). However, a noticeable decrease in the rotation velocity at the periphery is fairly rare even for massive galaxies. In the cases where this decrease occurs, it may not always reflect the circular velocity profile.

The observed decline in the rotation curve may have several reasons:

(1) underestimation of the rotation velocity due to measurement errors, which can potentially be caused by large-scale non-circular gas motions or disk plane curvature at the galaxy's periphery. The first factor usually manifests itself as an asymmetry in the rotation curve and becomes apparent in the analysis of a two-dimensional velocity field. The disk curvature, as a rule, is considered when using models in which the disk is divided into concentric rings ("tilted rings model" [7]) when processing the galaxy's velocity fields;

(2) a massive disk concentrated toward the center, which is responsible for high rotation velocity at $R \sim 2R_d$ ($R_d$ is the radial disk scale), where the disk contribution to the rotation curve is maximum;

(3) a massive dense bulge, which may be responsible for the disk's fast rotation in the galaxy's center, and, as a result, a decrease in the velocity at large $R$;

(4) a halo with low mass or weak concentration toward the center, which does not make a large contribution to the rotation curve within the measurement region.

The mass distribution modeling in galaxies with declining rotation curves allows one not only to estimate the masses of the main galaxy components within the standard Newtonian approach, but also to possibly investigate the application of non-Newtonian gravity models without dark matter, in which the asymptotic rotation velocity at large distances from the center $R$ corresponding to very low accelerations should approach a constant.

The most developed among alternative approaches is the modified Newtonian dynamics (MOND) model, which proposes an alternative to dark matter in order to explain the plateau of rotation curves at the galaxy periphery [8]. A detailed review of the MOND theory and its application to observations can be found, i.e., in [9]. MOND proposes the kind of acceleration in the gravitational field, $a_M(R)$, that asymptotically coincides with the classical Newtonian law $a_M = a_N$ at sufficiently large accelerations $a_M \gg a_0 \sim 10^{-8}$ cm/s$^2$. However, with decreasing acceleration, it increasingly deviates from the Newtonian value, so at $a_M \ll a_0$, the acceleration in the gravitating sphere field decreases as $R^{-1}$ instead of $R^{-2}$, which ensures a constant circular velocity given the fixed matter mass. In this case, parameter $a_0$ is universal.

The most compelling argument in favor of MOND is that it provides a simple explanation for the dependence between the mass of baryonic matter $M_{\rm bar}$ in galaxies and the outer region's rotation velocities of the disks $V_{\rm flat}$ (the so-called baryonic Tully–Fisher (BTF) relation); according to observations, this dependence has a surprisingly small scatter of points along the $M_{\rm bar} \sim V^4$ relation predicted by MOND [10, 11]. The universal character of the radial variation in the dispersion of stellar velocities in elliptical galaxies (according to MANGA) also does not contradict MOND [12].

The possibility of using MOND to interpret the galaxy's rotation curves has been considered in a number of studies, but the results remain inconsistent. In most cases, the use of an additional parameter does make it possible to interpret the galaxy's rotation curves without introducing a dark halo as successfully as if the halo was present (see, e.g., [13]). However, some authors question the universality of the parameter $a_0$ in the individual galaxy models, such as ESO138−G014 [14] or NGC 3109 [15]. It was shown [16] that, for dwarf and LSB galaxies, MOND adequately explains the observed rotation curve shape in 3/4th of cases.

MOND can be tested by the capability to explain rotation curves with a decreasing rotation velocity of the outer regions, as well as by comparing the estimates of stellar disk masses in this model with the values derived from photometric estimates.

The aim of this study is as follows:

(1) to search for specific features in the galaxies with declining rotation curves in comparison with galaxies with flat rotation curves;

(2) to verify how adequately declining rotation curves of galaxies are described in the MOND theory as compared to the classical approach, how the universal character of the constant $a_0$ is maintained, and how the mass disk estimates are consistent with their photometric characteristics.

## 2. GALAXY SAMPLES

### 2.1. Galaxy Sample with Declining Rotation Curves

The sample includes 22 spiral galaxies selected from the literature due to the following criteria:

(1) The galaxies do not belong to closely interacting systems (are not listed in the Arp and VV catalogs) and have no close neighbors of comparable luminosity.

(2) The measured rotation curve is not shorter than half the photometric radius $R_{25}/2$.

(3) The rotation velocity decrease after passing the maximum within the region covered by the existing rotation curve is at least 10–15%.

(4) The measured radial profile of the rotation velocity is symmetrical with respect to the center.

The perinuclear maximum of the rotation curve within ~1–1.5 kpc from the center, which is observed in some galaxies, was not considered; its cause is different than that of the farther maximum and is associated with the central region dynamics, which is primarily due to the existence of a dense bulge and/or contrasting bar.

The sample is certainly not exhaustive and contains few relatively close galaxies, which potentially now have the most reliably measured declining rotation curves. The references to the original sources on the rotation curves are given in Table 1.

One of the sample galaxies (NGC 753) is part of a cluster, and some of the galaxies belong to pairs or groups, but there are no close pairs among them. The sample includes NGC 3031, which is the main member of a group that shows interaction signs between the galaxies in the HI line. However, these are still considered, since NGC 3031 is sufficiently isolated from galaxies of comparable luminosity and has a well-studied symmetric rotation curve (see [17] and references to earlier studies).

The galactic photometric data and the flux estimates in the HI line were taken from the HyperLeda database[1] [32]. The distances to the galaxies were assumed to be the same as in the studies from which their rotation curves were taken. The distances in all galaxies outside the Local Group roughly correspond to the Hubble constant $H_0 = 75$ km/s/Mpc. Table 1 shows basic information about the galaxies, including distance, absolute magnitude, exponential disk scale $R_d$, and data from which the rotation curve was obtained (spectral line, as well as one-dimensional or two-dimensional velocity distribution).

---

[1] http://leda.univ-lyon1.fr

**Table 1.** Galaxy samples with declining rotation curves

| Galaxy | $d$, Mpc | $M_B$, mag | Lines | References | $R_d$, kpc | Reference |
|---|---|---|---|---|---|---|
| (1) | (2) | (3) | (4) | (5) | (6) | (7) |
| NGC 157 (SABb) | 20.9 | −21.2 | $H_\alpha$, H I (2D) | [18, 19] | 1.6 | [20] |
| NGC 224 (Sb) | 0.79 | −21.2 | H I (2D) | [21] | 4.5 | [21] |
| NGC 512 (Sab) | 68.5 | −21.3 | $H_\alpha$, [N II] (1D) | [21] | 2.9 | [21] |
| NGC 582 (SBb) | 61.6 | −21 | $H_\alpha$, [N II] (1D) | [21] | 3.4 | [21] |
| NGC 753 (SABc) | 66.8 | −21.8 | $H_\alpha$ (2D) | [22] | 4.6 | [23] |
| NGC 1642 (Sc) | 62.1 | −21.1 | $H_\alpha$, H I (2D) | [24] | 3.5 | [24] |
| NGC 2599 (Sa) | 68.4 | −21.3 | $H_\alpha$, H I (2D) | [24] | 9.0 | [24] |
| NGC 2841 (SBb) | 14.1 | −21.2 | H I (2D) | [17] | 4.6 | [17] |
| NGC 2903 (Sbc) | 8.9 | −20.9 | H I (2D) | [17] | 2.8 | [17] |
| NGC 3031 (Sab) | 3.6 | −20.7 | H I (2D) | [17] | 1.4 | [25] |
| NGC 3521 (SABb) | 10.7 | −20.9 | H I (2D) | [17] | 1.2 | [17] |
| NGC 3719 (Sbc) | 78.6 | −21.2 | FP $H_\alpha$ (2D) | [22] | 3.6 | * |
| NGC 3893 (SABc) | 15.5 | −20.7 | FP $H_\alpha$ (2D) | [26] | 2.1 | [26] |
| NGC 3992 (Sbc) | 18.6 | −21.3 | H I (2D) | [27] | 2.3 | [28] |
| NGC 4138 (S0-a) | 20.7 | −19.4 | N II, H I (2D) | [29] | 1.4 | [20] |
| NGC 4725 (SABa) | 11.9 | −20.7 | H I (1D) | [30] | 4.3 | [30] |
| NGC 4736 (SABa) | 4.7 | −19.8 | H I (2D) | [17] | 1.5 | [17] |
| NGC 5055 (Sbc) | 10.1 | −21.1 | H I (2D) | [17] | 3.6 | [17] |
| NGC 5297 (Sc) | 35 | −21.4 | FP $H_\alpha$ | [22] | 5.74 | [31] |
| NGC 7793 (Scd) | 3.9 | −18.7 | H I (2D) | [17] | 1.3 | [17] |
| UGC 10692 (Sb) | 130 | −21.5 | H I (1D) | [21] | 8.6 | [21] |
| UGC 10981 (Sbc) | 151 | −21.9 | $H_\alpha$, [N II] (1D) | [21] | 5.1 | [21] |

The columns show (1) the name and type of the galaxy; (2) accepted distance, Mpc; (3) absolute magnitude $M_B$, mag; (4) spectral lines from which the rotation curve was plotted: slit (1D) or two-dimensional (2D) spectroscopy; (5) rotation curve source; (6) disk scale $R_d$, kpc; and (7) disk scale source. The asterisk (*) indicates the data from the present study.

The observed rotation curves for the galaxies in our sample are shown in Fig. 1.

The galaxies under study include galaxies both with and without a bar; specifically, out of 22 galaxies, two galaxies have strong bars (SB type) and 14 galaxies show no apparent signs of a bar. Therefore, bars are not the cause of the observed decline in rotation curves. The table contains galaxies of early (e.g., NGC 4138 and NGC 2599) and late morphological types (e.g., NGC 5297 and NGC 7793), galaxies with flocculent arms (NGC 2841, NGC 3521, NGC 5055) and those with a Grand Design structure (NGC 3031, NGC 2903, NGC 3992, NGC 157). It can be concluded that galaxies with declining rotation curves have no significant morphological peculiarities.

### 2.2. Galaxy Subsample with the Longest Rotation Curves

In order to decompose the rotation curve (see Section 4), galaxies from Table 1 was used. Its rotation curve, according to available observations in the HI line, can be traced beyond the optical radius. Galaxies from Table 2 show that rotation curves were obtained using the "tilted rings model", i.e., the variation in the disk inclination with the distance from the center is

**Table 2.** Galaxy subsamples with long rotation curves

| Galaxy | $R_{25}$, kpc | $R_{lim}$ | $\Delta V/V_{max}$ | References |
|---|---|---|---|---|
| (1) | (2) | (3) | (4) | (5) |
| NGC 157 | 11.3 | $2R_{25}$ | 46 | [18] |
| NGC 2841 | 13.8 | $2R_{25}$ | 16 | [33] |
| NGC 2903 | 18.1 | $1.4R_{25}$ | 14 | [34] |
| NGC 3031 | 12.0 | $1.2R_{25}$ | 24 | [3] |
| NGC 3521 | 13.2 | $2R_{25}$ | 17 | [33] |
| NGC 3992 | 24.5 | $1.1R_{25}$ | 11 | [27] |
| NGC 5055 | 17.3 | $2.2R_{25}$ | 18 | [33] |

The columns show (1) the galaxy's name; (2) optical radius $R_{25}$, kpc; (3) distance $R_{lim}$ at which the rotation velocity decreases by $\Delta V$; (4) ratio of $\Delta V$ to the maximum rotation velocity $V_{max}$; and (5) references to studies on the HI density distribution.

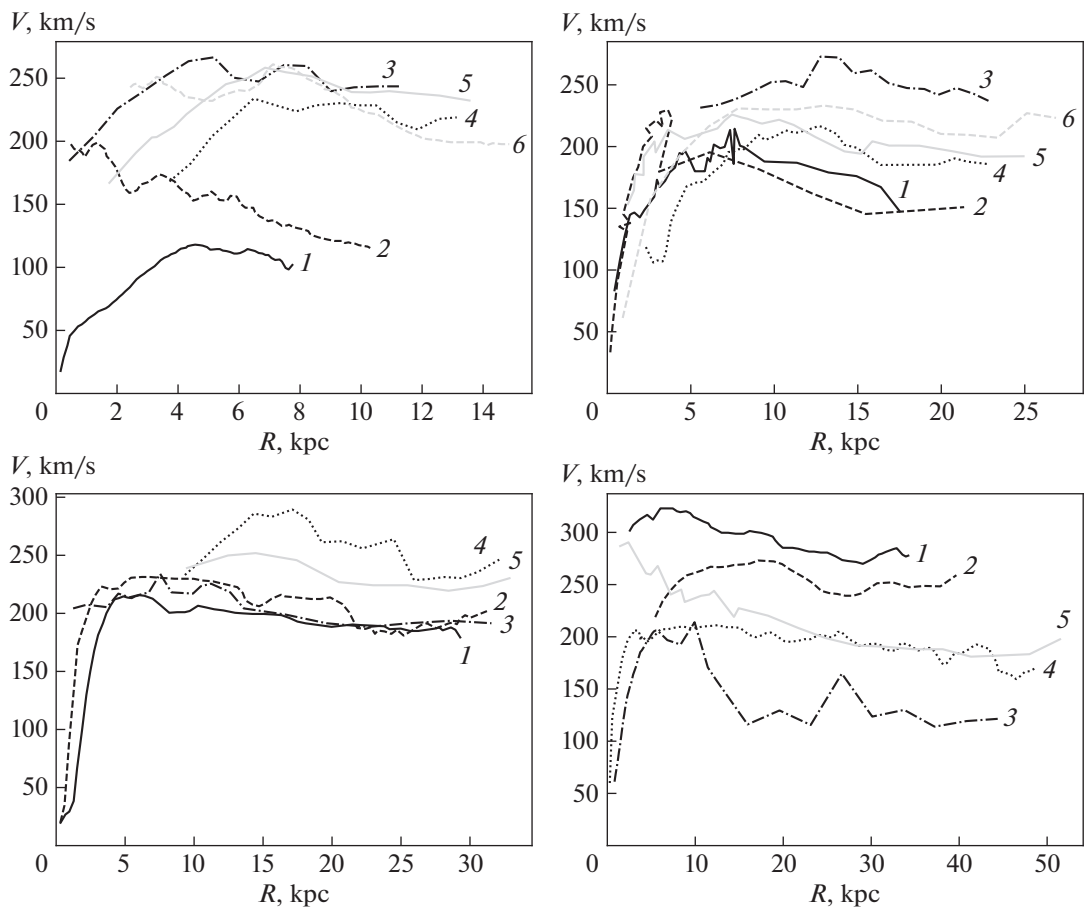

**Fig. 1.** Observed rotation curves of the galaxies from the sample. Top panel, left: NGC 7793 (*1*), NGC 4736 (*2*), NGC 512 (*3*), NGC 582 (*4*), NGC 3719 (*5*), NGC 3031 (*6*); right: NGC 3893 (*1*), NGC 4138 (*2*), UGC 10981 (*3*), NGC 5297 (*4*), NGC 753 (*5*), NGC 4725 (*6*). Bottom panel, left: NGC 2903 (*1*), NGC 3521 (*2*), NGC 1642 (*3*), UGC 10692 (*4*), NGC 224 (*5*); right: NGC 2841 (*1*), NGC 3992 (*2*), NGC 157 (*3*), NGC 5055 (*4*), NGC 2599 (*5*).

already considered and should not be the cause for the rotation curve decline. The table shows the photometric radius $R_{25}$, the distance $R_{lim}$ at which the rotation velocity decreases by $\Delta V$, and the ratio of $\Delta V$ to the maximum rotation velocity $V_{max}$. All rotation curves of the subsample galaxies described above are shown in Fig. 1.

Below are brief summaries on individual galaxies of this subsample.

**NGC 157.** The distance from NGC 157 to the nearest galaxy of comparable luminosity is at least 1.3 Mpc [18], i.e., this galaxy can be considered isolated. The galaxy's spiral pattern is the Elmegreen class 12 [35], i.e., the galaxy has distinct symmetrical spiral arms. However, the regular spiral structure extends only to a radius of approximately 1′ (~6 kpc); furthermore, the structure becomes flocculent or breaks up into multiple arms [19]. HI observations show a disk curvature obtained from kinematic data and a significant decrease in the rotation velocity immediately after the optical radius $R_{25}$, which suggests low mass or low concentration of the dark halo [18].

**NGC 2841.** This close massive galaxy with flocculent spirals has a bright classic bulge [36]. The galaxy is isolated and shows no traces of past interactions [37].

**NGC 2903.** This is a galaxy with an active star formation, which can also be considered isolated [38]. It is rich in gas and has a stellar bar and two symmetrical long spiral arms.

**NGC 3031.** This is a close galaxy with distinct symmetrical spiral arms (type 12 according to the Elmegreen classification [35]). The hump in the rotation curve is observed at a distance of ~7.5 kpc from the center [17]. Strong spiral arms and related noncircular gas motions are responsible for local features on the rotation curve; however, the galaxy's outer regions clearly rotate slower than the inner ones. This galaxy forms an interacting system with M82 and NGC 3077 (see, e.g., [39]).

**NGC 3521.** According to the Elmegreen spiral arm classification, NGC 3521 is a type 3 galaxy [35], i.e., it has flocculent spirals. This galaxy has a very large and bright classic bulge with star formation zones [36]. It has no close neighbors of comparable luminosity.

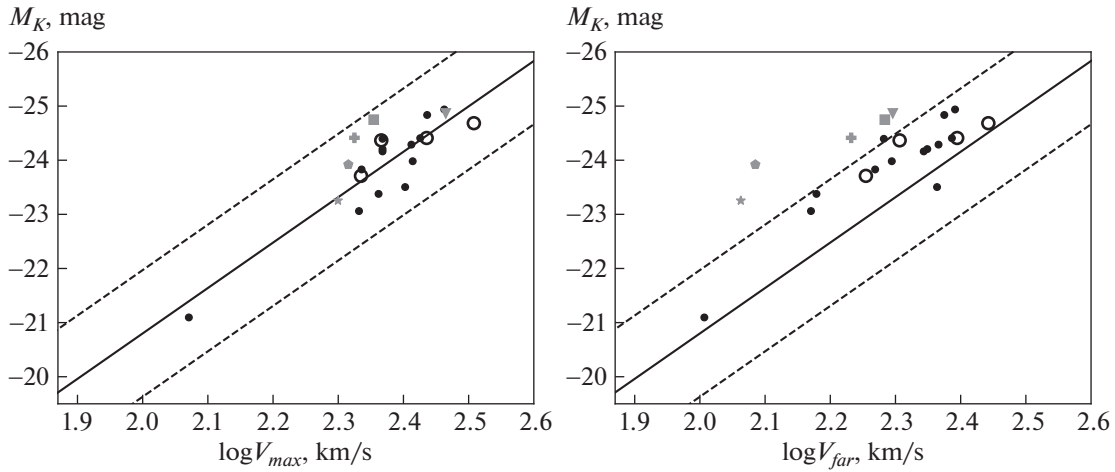

**Fig. 2.** Tully–Fisher (TF) relations in the $K$ band for the maximum rotation velocity $V_{max}$ (left) and velocity at the periphery $V_{far}$ (right) for the galaxies with declining rotation curves. The straight lines show the TF relations for the GHASP galaxies; the dashed lines correspond to the zero-point error ($1\sigma$) of the relation according to [42]. The galaxies that exceed $1\sigma$ are indicated in the figure on the right: the triangle is NGC 2599, the square is NGC 753, the cross is NGC 5055, the pentagon is NGC 157, and the asterisk is NGC 4736; the galaxies from the subsample with the longest rotation curves are marked with circles (see section 2.2).

**NGC 3992.** This is one of the largest galaxies of the Ursa Major cluster. It has a strong bar and two symmetrical, tightly twisted spiral arms with branches and "spurs."

**NGC 5055.** This is a type 3 galaxy according to the Elmegreen' arms classification [35], i.e., it has a flocculent spiral pattern. The galaxy has a pseudo-bulge with an active star formation [36]. NGC 5055 has a broad neutral-hydrogen disk that is curved at the periphery and extends much further than the optical disk [40]. The galaxy belongs to the close group M51 [41]. The near-est galaxy of comparable luminosity (UGC 8313) is located at a distance of 70 kpc (in the picture plane) from NGC 5055.

## 3. COMMON GALAXY PROPERTIES WITH DECLINING ROTATION CURVES

### 3.1. Tully–Fisher Relation and Baryonic Tully–Fisher Relation

Figure 2 shows the positions of the galaxies on the TF relation, where the absolute magnitude $M_K$ is matched with the maximum velocity $V_{max}$ (Fig. 2, left) and the rotation velocity of the regions farthest from the center $V_{far}$ (Fig. 2, right). For comparison, straight lines show the linear dependence obtained in [42] for the galaxies of the GHASP survey. The galaxies of this survey were chosen as reference galaxies, since their rotation curves are determined with good angular and linear resolution (from observations in the $H_\alpha$ line) using a unified procedure and based on two-dimensional velocity fields. The latter allows for a more accurate measurement of the disk's orientation angles, and hence for the rotation velocity in the region of maximum or where it reaches a plateau. Among the 83 GHASP galaxies, only three show a decrease in the rotation curve (one of them is included in our sample).

The TF relation obtained in [42] for the GHASP galaxies with $M_K < -20$ has the form

$$M_K = (-4.02 \pm 1.17) - (8.39 \pm 0.52) \log V_{max}. \quad (1)$$

For our galaxy sample with declining rotation curves, the $M_K$ values were determined from integrated quantities $K_s$ or $K$ taken from the NED database. The circles indicate the galaxies with the longest rotation curves (see Section 2.2). The straight line in the figures is the dependence for the GHASP galaxies [42]. The dashed lines correspond to the zero point's error ($1\sigma$) of this dependence (according to [42]). Almost all GHASP galaxies with symmetrical rotation curves lie inside the corridor of values bounded by these lines.

As follows from the above diagrams, galaxies with declining rotation curves agree better with dependence (1) if their rotation velocity is characterized with $V_{max}$ (the median of the velocity deviation for $\log V_{max}$ is 0.04; for $\log V_{far}$, it is 0.09); this effect is more pronounced for the subsample galaxies with long rotation curves. Thus, the maximum rotation velocity is quite normal for the luminosity of galaxies with declining rotation curves. The decrease in the rotation curve is apparently associated with a reduced (for a given luminosity of the galaxy) rotation velocity of the disk's outer regions, and not with an overvalued rotation velocity of the central regions. In this case, the decline of the rotation curve at large $R$ should be not due to a disk or bulge being more concentrated toward the center, but rather because of the less concentrated or less massive halo.

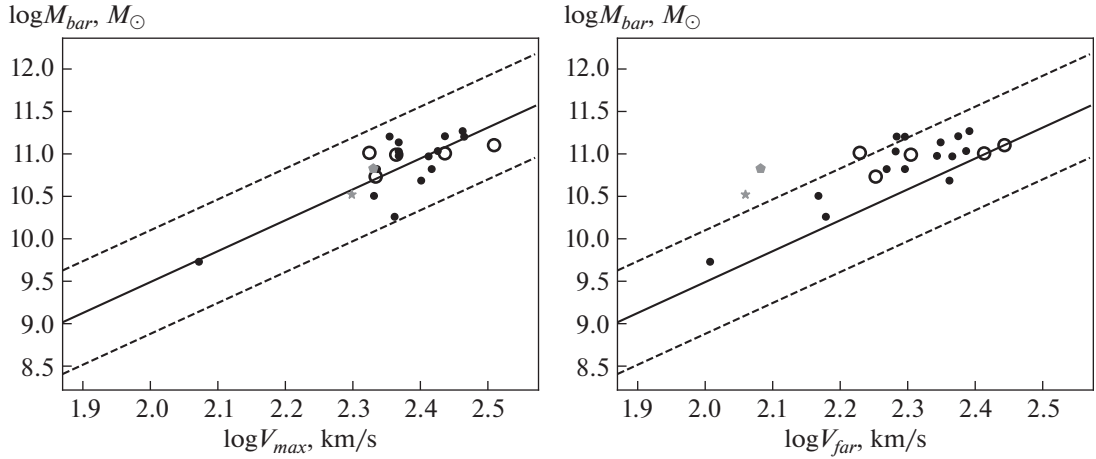

**Fig. 3.** Galaxy positions on the baryonic Tully–Fisher (BTF) relation for the maximum (left) and peripheral (right) rotation velocities. The notations are the same as in Fig. 2.

Note that a different conclusion was obtained in [2], where the TF relationships for the galaxies of the UMa cluster were analyzed: the galaxies whose rotation velocities pass through the maximum turn out to be on the same dependence with galaxies with a flat rotation curve, if the velocity $V_{\rm far}$ at large $R$ is considered. A similar inference was obtained later in [11] for a sample of 32 galaxies with rotation curves taken from various sources. The discrepancy with our conclusion is caused by the difference in the slopes and zero-points of the compared TF relations, as well as by the small number of galaxies with a radial decrease in rotation velocity in the cited works (5 galaxies in [2] and 9 galaxies in [11]).

The BTF relationship between the baryonic mass and the disk rotation velocity has a more deep physical sense. As in the case of the TF relation, we used the sequence obtained in [42] for the GHASP galaxies as the reference BTF relation:

$$\log M_{\rm bar} = (2.21 \pm 0.61) + (3.64 \pm 0.28)\log V, \quad (2)$$

where $M_{\rm bar}$ is the galaxy's total baryonic mass, and $V$ is the rotation velocity.

The baryonic mass for the galaxies in our sample was determined as the sum of the stellar mass and the HI mass (multiplied by 1.4 to take into account heavier elements), which was calculated from the parameter $m21$ (from HyperLeda):

$$m21 = -2.5\log f + 17.40, \quad (3)$$

where $f$ is the integral radiation flux in the HI line in Jy km/s.

The median ratio of gas mass to baryonic mass is ~0.2 for the sample as a whole and ~0.1 for the galaxies with the longest rotation curves; the maximum value of this ratio is 0.34. Since the gas mass for these galaxies is only a small fraction of the baryonic mass, errors in its determination have little effect on the position of the galaxies on the BTF diagram.

The stellar mass of the galaxies $M_*$ was estimated by the absolute magnitude $M_K$; the $M/L_K$ ratio for comparison with the galaxies of the GHASP reference sample was taken equal to 0.8, same as in [42]. This value follows from the de Jong model [43] for a stellar population with an age of 12 Gyr with solar metallicity, constant star-formation rate, and the Salpeter initial mass function. It is important to note that this $M/L_K$ value is apparently overestimated in most galaxies with active star formation (see, e.g., [11, 44]). In Fig. 3, the straight line indicates dependence (2) for the GHASP galaxies; the dashed lines show the corridor of 1σ, which reflects the accuracy of determination of the zero-point of relationship [42]. The circles indicate the galaxies from the subsample with the longest rotation curves.

Note that on the right graph where $V_{\rm far}$ is used, the point scatter is larger as compared to the left graph; this reflects a wider range of $V_{\rm far}$ in comparison with $V_{\rm max}$.

The right panel of Fig. 3, where the velocity of the galaxy's outer regions is matched with the baryonic mass, shows a systematic shift with respect to the dependence for the GHASP galaxies (average deviation of $\log V_{\rm far} = 0.09$), which is approximately equal to the shift in the TF relation. From both the TF and BTF relations, it follows that at least three galaxies (NG 157, NGC 4736, and NGC 5055) have an abnormally low rotation velocity or $V_{far}$ for a given mass luminosity.

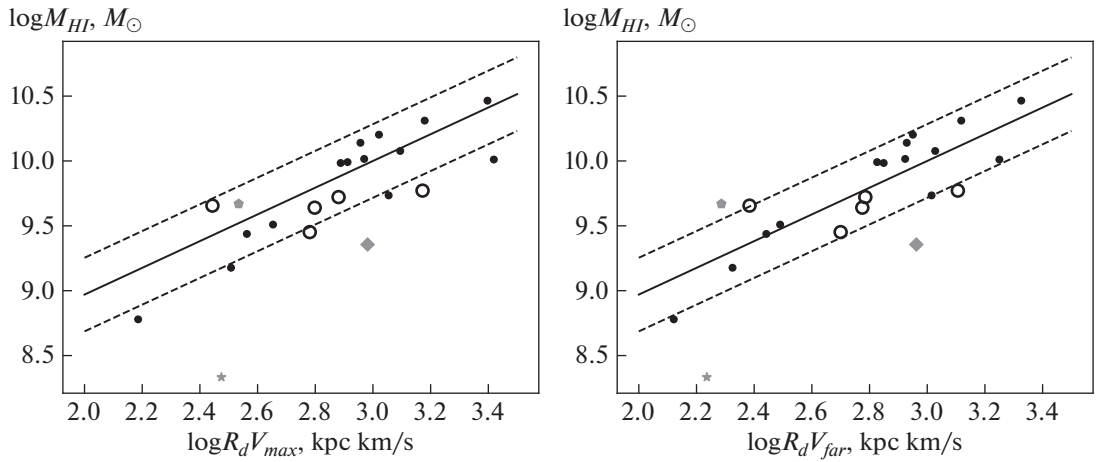

**Fig. 4.** Galaxy positions under study in the diagrams that relate the total HI mass to the parameters characterizing the specific angular momentum. The straight lines refer to isolated late-type galaxies used for comparison according to [45]. The dashed lines show the scatter in 1σ for the reference galaxies. The figure on the right shows the galaxies that deviate from the reference dependence the most: the asterisk is NGC 4736, the diamond is NGC 4725, and the pentagon is NGC 157; the circles indicate the galaxies from the subsample with the longest rotation curves.

### 3.2. Neutral Hydrogen Content

Following earlier studies, a close correlation between the total hydrogen mass $M_{HI}$ and the specific moment of the disk rotation is known, which is assumed to be proportional to the product of the rotation velocity and the diameter $D_{25}$ or radial disk scale $R_d$ (see, e.g., [45] and references therein). Since the photometric diameter, unlike the radial disk scale, depends on the disk surface brightness, the relations of $\log M_{HI}$ to $R_d V_{max}$ and $R_d V_{far}$ were considered. In Fig. 4, they are compared with the relation

$$\log M_{HI} = 1.03 \log(R_d V) + 6.91, \quad (4)$$

obtained in [45] for isolated late-type spiral galaxies with a moderate tilt angle of the disk (solid line). The $M_{HI}$ mass is expressed in solar masses, $R_d$ in kpc, and $V$ in km/s. The circles indicate the subsample galaxies with the longest rotation curves.

The median deviations of the sample points from dependence (4) for both $V_{max}$ and $V_{far}$ are smaller than the scatter of 1σ relative to the solid straight line. Therefore, the considered galaxies with declining rotation curves do not noticeably differ in the neutral hydrogen content from the galaxies for which expression (4) was obtained. A similar plotted relation using optical diameter $D_{25}$ instead of $R_d$ (not shown here) demonstrates a similar result. It is important to note that two of the galaxies under study (NGC 4725 and NGC 4736), which are of the Sa type, have a HI deficit; this is quite expected when comparing the two, since the reference dependence is related to late-type galaxies.

## 4. DECOMPOSITION OF LONG ROTATION CURVES IN THE NEWTONIAN GRAVITY AND MOND MODELS

In the previous section, only rough estimates of the galaxy's total baryonic mass were used. To model the selected galaxy's rotation curve and decompose it into constituents associated with various components of the galaxies, it is necessary to know the distribution of the stellar and gas mass along the radius. While the distribution of the observed gas mass is inferred from the HI observations, the stellar population mass is a more complicated task to consider. The mass of stellar population $M_*$ was estimated in two ways: first, through the luminosity in the near-IR range and the $M/L$ ratio corresponding to the galaxy's color index; second, by modeling the rotation curve in the model of the maximum disk mass with free choice of $M/L$, which gives an upper limit estimate of $M_*$.

### 4.1. Newtonian Decomposition of the Rotation Curve

The rotation curve was represented by the sum of four components: disk, bulge, gas layer, and halo. The disk and gas layer were considered thin, while the bulge and halo were proposed to be spherical. It was assumed that the radial distribution of the stellar disk surface density follows the photometric profile in the near-IR range. To distinguish the disk's brightness profile, it was conditionally assumed that the disk's brightness prevails over the bulge's brightness starting from the radius where the radial brightness profile expressed in a logarithmic scale sharply rises to the center. The transition from the brightness distribution of the galactic disk to the density distribution was carried out by multiplying the brightness in solar units $[L_\odot/\text{pc}^2]$ by the mass-to-luminosity ratio $M/L$, which was generally considered to be dependent on the color index (see below). For the disk's distant regions with no direct measurements of the photometric profile, the latter was approximated by an exponential law with a scale corresponding to the adjacent regions closer to the center. It is important to note that the disk's contribution to the rotation curve on the galaxy's periphery, in the presence of a massive halo, is usually small; therefore, the error associated with such extrapolation is low.

The surface brightness in units of [mag/arcsec$^2$] was converted to solar units $[L_\odot/\text{pc}^2]$ through the relation

$$\log \mu\left[\frac{L_\odot}{\text{pc}^2}\right] = 0.4\left(21.57 - \mu\left[\frac{\text{mag}}{\text{arcsec}^2}\right] + M_{\odot,K}\right), \quad (5)$$

where the absolute magnitude of the Sun in the $K$ band, $M_{\odot,K}$, was taken as 3.28 [46].

The mass-to-luminosity ratio at 3.6 μm was estimated using the dependence of $M/L_{[3.6]}$ versus $B - V$ proposed by McGaugh et al. [47] for the model [48]:

$$\log M/L_{[3.6]} = -0.861 + 0.849(B - V). \quad (6)$$

The values of $M/L_{[3.6]}$ and $M/L_K$ are closely related, since both belong to the near-IR range. Following [49], we used the linear relation

$$M/L_{[3.6]} = 0.92 M/L_K - 0.05. \quad (7)$$

For the galaxies under study, the obtained $M/L_K$ ratios lie in the range between 0.44 and 0.76 solar units. It should be noted that mass-to-luminosity ratios depend on the stellar evolution models and on parameters that are difficult to consider, such as the initial mass function and the star's chemical composition. Therefore, various models give noticeably different values; however, even for the "reddest" galaxies with a solar abundance, they rarely exceed 0.8, which is consistent with independent dynamic estimates of the disk masses (see the discussion in [4, 11]).

The rotation curve's decomposition was carried out using the GRVolSu software developed by D.S. Mukhatov (Volgograd State University), which simulates the galaxy's rotation curves consisting of various components with a given mass distribution profile according to Newtonian mechanics. The rotation curve was compared with a model one that included a bulge (King's model with the mass determined from the photometric profile), thin stellar disk, thin gas disk, and halo (NFW profile).

The parameters of the components were selected to be fitted with the observed curve of rotation for the maximum disk model (see f.e. [50]) with a minimum error. For galaxies where the bulge (pseudo-bulge) has a relatively small mass, the results weakly depend on the bulge's adopted model (which justifies using a rather crude approximation for bulge parameters): the bulge defines the rotation curve shape only in the inner disc region and has little effect on the velocity of the rotation curve's declining part.

When using the isothermal halo profile instead of the NFW profile, same as when considering the finite disk thickness ($2z_0/D \sim 0.2$, where $z_0$ and $D$ are the half-thickness and diameter of the disk, respectively), verification showed that the results qualitatively remain the same.

### 4.2. Rotation Curve Plotting in the MOND Model

According to the MOND theory, the acceleration of a body in a circular orbit, $a_M$, at low accelerations exceeds the acceleration $a_N$ expected in the Newtonian theory of gravity. The difference between these accelerations is negligible at $a_N \gg a_0$ and significant when $a_N \approx a_0$ or smaller, where $a_0$ is a universal physical parameter. Interpretation of the BTF relation in the context of MOND leads to the estimate $a_0 = (1.3 \pm 0.3) \times 10^{-8}$ cm/s$^2$ [10]. The transition between these limiting cases can be described by a simple expression [30]: $a_N = a_M \mu(x)$, where $\mu(x) = x/\sqrt{1 + x^2}$ and $x = (a_M/a_0)$. At low accelerations, $a_M = \sqrt{a_N a_0}$.

The rotation curve in the MOND model was constructed using the Newtonian model of the galaxy with the halo removed, i.e., the radial density profiles of baryonic components taken from photometry were preserved, but the mass-to-luminosity ratio of the stellar population was considered as a free parameter. As the next step, the Newtonian acceleration $a_N$ associated with the baryonic components (bulge and stellar disk) was calculated for every galaxy using a series of $M/L_K$ ratios: $a_N = V_N^2/R$ where $V_N$ is the rotation velocity expected in rotation velocity expected in the absence of a halo. Then the correspong values of $a_m$ were determined using the above given ratios. Thus, the model of the galaxy rotation curve corresponding to MOND (see Figs. 5–11) for the selected $M/L_K$ was constructed. The value $a_0$ considered as a free parameter; it's selection was based on the best agreement (through visual inspection) of the observed rotation curve for the galaxy's outer regions with the model curve.

Note that in MOND the accelerations that are compared with $a_0$ include those associated with both internal gravitational forces in the system and the external fields (with an obvious violation of the strict equivalence principle, as discussed in [9]). However, in our case, thi circumstance is irrelevant: the galaxies selected for modeling have no massive neighbors capable of creating a predominant external acceleration.

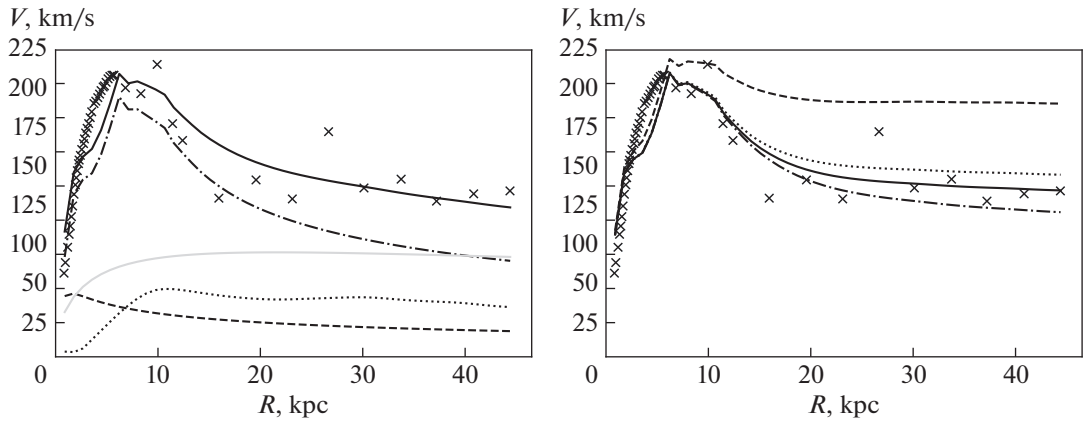

**Fig. 5.** NGC 157. Left: rotation curve decomposition in the Newtonian maximum disk model. The crosses show the observed rotation curve, the solid line shows the model rotation curve, and other lines show various component profiles: bulge (dotted line), disk (dash-dotted line), gas (dashed line), halo (thin solid line). Right: comparison with the MOND models; the crosses show the observed rotation curve; the dashed, solid, dotted, and dash-dotted lines show the rotation curves for the models with $a_0 = 1.2, 0.2, 0.3,$ and $0.1 \times 10^{-8}$ cm/s$^2$, respectively.

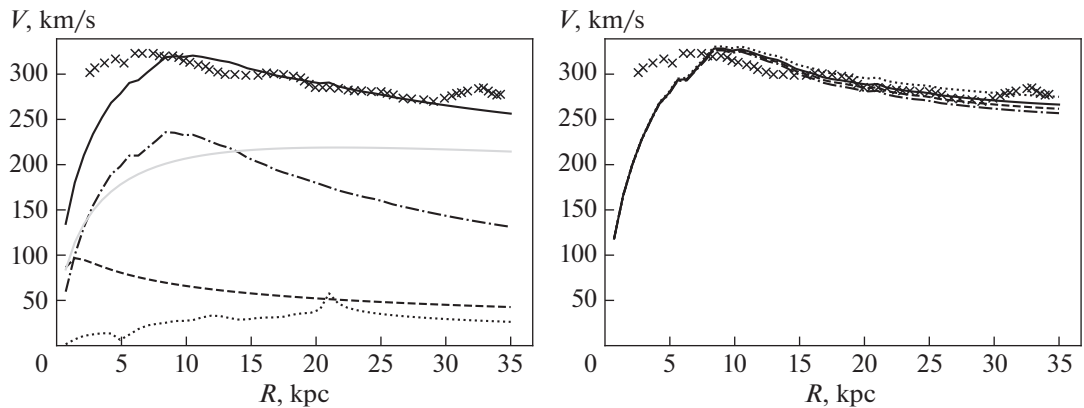

**Fig. 6.** NGC 2841. Left: rotation curve decomposition in the Newtonian maximum disk model; the notations are the same as in Fig. 5. Right: comparison with the MOND models; the observed rotation curve is indicated by crosses; the dashed, solid, dotted, and dash-dotted lines show the dependences for the MOND model at $a_0 = 1.2, 1.3, 1.5,$ and $1.1 \times 10^{-8}$ cm/s$^2$, respectively.

### 4.3. Results and Discussion

The results of rotation curve decomposition in the Newtonian model of gravity and the rotation curve construction in the MOND model for seven galaxies included in the subsample of galaxies with the longest rotation curves are shown in Figs. 5–11 and in Table 3. The mass-to-luminosity ratios for the Newtonian model ($M/L_K^N$) and for MOND ($M/L_K^M$) are denoted with superscripts $N$ and $M$, respectively (indicated by photometric band).

Below are brief summaries on these seven galaxies:

**NGC 157.** The model constructed under the assumption of Newtonian gravity explains the observed rotation curve near the maximum and at the periphery. The resulting $M/L_K^N$ ratio differs from the value based on photometric estimates by ~30%. In MOND, the best fit to the observed rotation curve is achieved at a very low value $a_0 = 0.2 \times 10^{-8}$ cm/s$^2$. However, the $M/L_K^M$ ratio required to explain the rotation curve turns out to be significantly higher than the value that corresponds to the stellar population color, and the disk's mass significantly exceeds the mass of the maximum disk in the Newtonian model.

**NGC 2841.** The observed rotation curve, except for the galaxy's central region, is satisfactorily described by the model with Newtonian gravity, although the ratio $M/L_K^N = 0.96$ in the maximum disk model is higher than the values obtained from stellar models, which generally do not exceed $0.60$–$0.7$ in the $K$ or 3.6 μm band for spiral galaxies [11, 17, 47]).

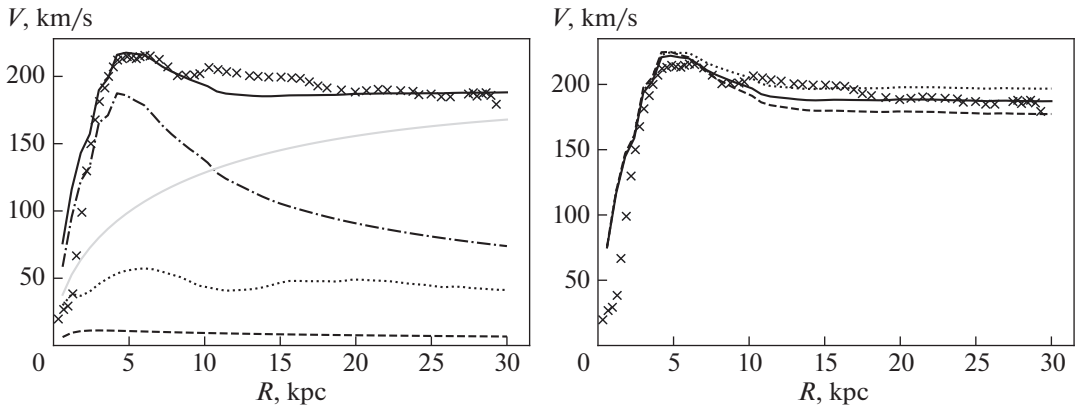

**Fig. 7.** NGC 2903. Left: rotation curve decomposition in the Newtonian maximum disk model; the notations are the same as in Fig. 5; Right: comparison with the MOND models; the observed rotation curve is indicated by crosses; the dashed, solid, dotted, and dash-dotted lines show the dependences for the MOND model at $a_0 = 1.2$, $1.6$, and $2.0 \times 10^{-8}$ cm/s$^2$, respectively.

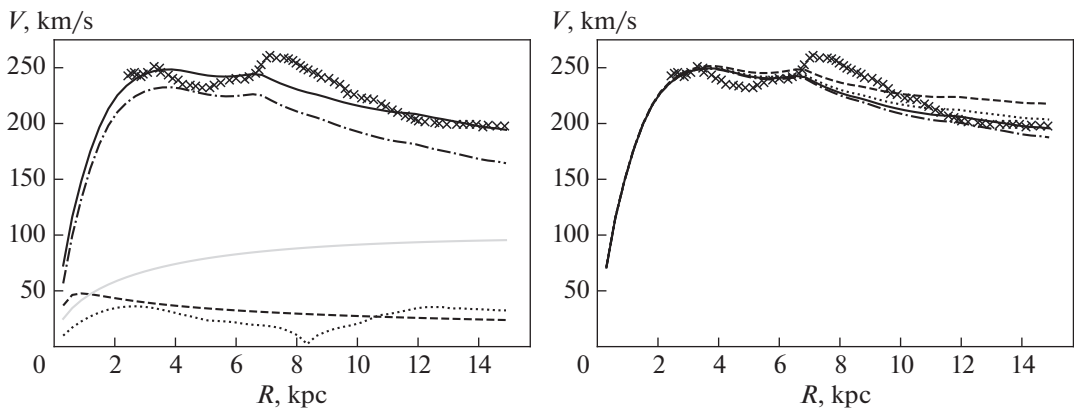

**Fig. 8.** NGC 3031. Left: rotation curve decomposition in the Newtonian maximum disk model; the notations are the same as in Fig. 5. Right: comparison with the MOND models; the observed rotation curve is indicated by crosses; the dashed, solid, dotted, and dash-dotted lines show the dependences for the MOND model at $a_0 = 1.2$, $0.6$, $0.8$, and $0.4 \times 10^{-8}$ cm/s$^2$, respectively.

The simulations of the NGC 2841 rotation curve in the MOND model (Fig. 6) agree with the observations almost as closely as in the maximum disk model. If $a_0$ is chosen so that the observed rotation curve at large $R$ is described best, there will be a noticeable discrepancy in the maximum region. If the parameter $a_0$ is reduced to describe the maximum region, the periphery's rotation curve will be slightly lower than the observed one. However, the disk's mass for the adopted values of $a_0$ is approximately two times the stellar population's mass obtained from photometric data, and is ~50% higher than the mass for the maximum disk model in the context of Newtonian gravity.

**NGC 2903.** If the $M/L_{[3.6]}^N$ ratio value in the decomposition of the observed rotation curve is fixed and considered as a free parameter, its optimal value will be close to 1.4, which is more than two times higher than the values for old stellar populations based on the stellar population models [11, 47]. However, this is close to an abnormally high $M/L_{3.6}^N = 1.23$ obtained for this galaxy's outer regions in the dynamic model with the mass-to-luminosity ratio variable along the radius from [17], which indicates photometric-type peculiarities of its stellar population. Thus, in both models (Newtonian and MOND) of the present study, the $M/L_{[3.6]}^N$ ratio varying along the radius obtained by de Blok et al. [17] was used.

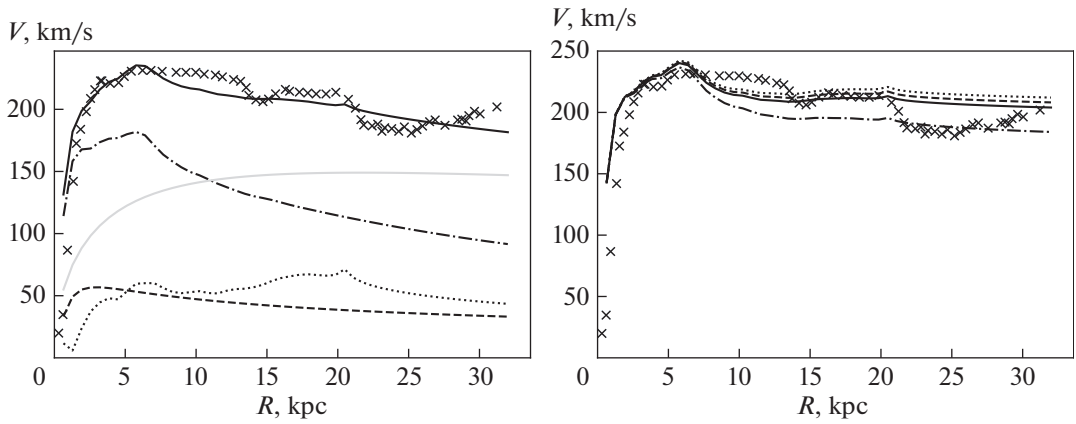

**Fig. 9.** NGC 3521. Left: rotation curve decomposition in the Newtonian maximum disk model; the notations are the same as in Fig. 5. Right: comparison with MOND models; the observed rotation curve is indicated by crosses; the dashed, solid, dotted, and dash-dotted lines show the dependences for the MOND model at $a_0$ = 1.2, 1.1, 1.3, and 0.7 × 10$^{-8}$ cm/s$^2$, respectively.

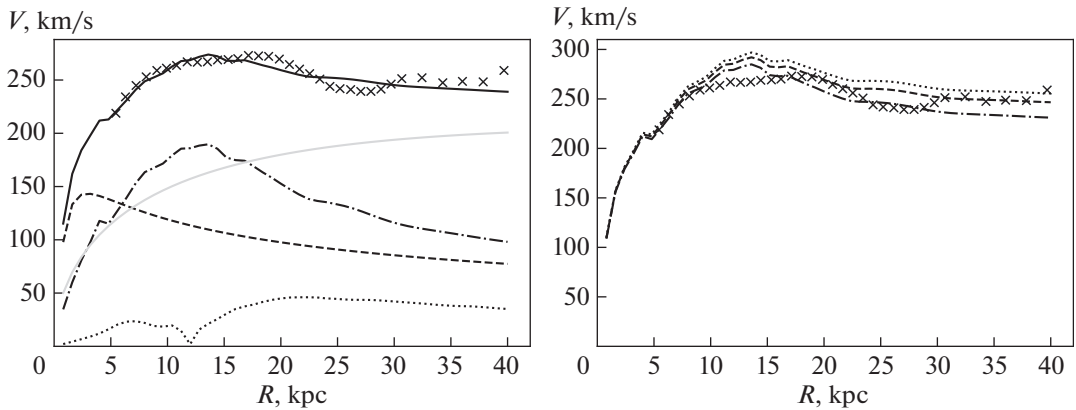

**Fig. 10.** NGC 3992. Left: rotation curve decomposition in the Newtonian maximum disk model; the notations are the same as in Fig. 5. Right: comparison with MOND models; the observed rotation curve is indicated by crosses; the dashed, solid, dotted, and dash-dotted lines show the dependences for the MOND model at $a_0$ = 1.2, 1.4, and 0.9 × 10$^{-8}$ cm/s$^2$, respectively.

Figure 7 shows that the rotation curve in the Newtonian model is well decomposed into components except for the 10–17 kpc region and the central region, where the observed velocities are lower than the model ones, although the bulge's contribution to the model rotation curve is small. The model rotation curve of NGC 2903 in MOND, when an appropriate $a_0$ value is chosen, also agrees with the rotation velocity profile in the galaxy's outer region, but poorly describes the maximum region, while the Newtonian model curve describes both the periphery and the maximum region well. Both models poorly explain the observed gas velocities at a radial distance of 10–17 kpc, where non-circular motions are apparently present. The value of the parameter $a_0$ found for MOND model is consistent with that adopted in this theory. However, the $M/L_K^M$ ratio in MOND is almost 3 times higher than the value based on photometry and is ~50% higher than the value in the maximum disk model.

**NGC 3031.** This galaxy's maximum disk model was calculated so that it describes the maximum clos- est to the galaxy's center, since the observed rotation curve in the region of the farther maximum at ~7.5 kpc is apparently distorted by non-circular gas motions in strong spiral arms [17]. Therefore, the constructed model explains well the rotation curve up to 4 kpc and after 11 kpc, while the photometric estimate of $M/L_K$ is consistent with the estimate obtained in the maximum disk model.

The MOND model best describes the observed rotation curve for the parameter values $a_0 \sim (0.6 \pm 0.2) \times 10^{-8}$ km/s$^2$, which is approximately half the expected value for MOND [10]. Nevertheless, the disk's mass in this model does not contradict the photometric estimate, as well as the estimate obtained for the maximum Newtonian disk.

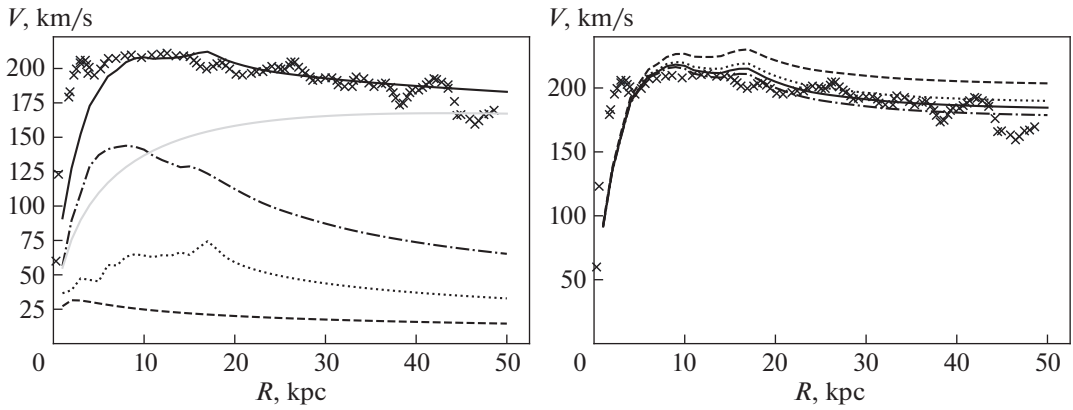

**Fig. 11.** NGC 5055. Left: rotation curve decomposition in the Newtonian maximum disk model; the notations are the same as in Fig. 5. Right: comparison with MOND models; the observed rotation curve is indicated by crosses; the dashed, solid, dotted, and dash-dotted lines show the dependences for the MOND model at $a_0 = 1.2, 0.8, 0.9$, and $0.7 \times 10^{-8}$ cm/s$^2$, respectively.

**NGC 3521.** Both the Newtonian and MOND models poorly describe the observed rotation curve of NGC 3521, which has a complex shape apparently related to local non-circular gas velocities. The Newtonian model well reproduces the 3–5 kpc region and the "smoothed" curve shape at large distances from the center. It is important to note that the 3.6 μm photometry and the color dependence of $M/L$ used by McGaugh and Schombert [47] predict an almost twice as dense disk for this galaxy than the rotation curve allows.

In the MOND model, the rotation curve's shape of NGC 3521 is poorly described. The values of the parameter $a_0$ reproducing the curve's general shape are close to $\sim 10^{-8}$ cm/s$^2$, which is consistent with the expected value in MOND. The $M/L_K^M$ ratio for MOND is slightly different (by ~20%) from the value based on photometric estimates.

**NGC 3992.** This galaxy's rotation curve is well described by the Newtonian model. The $M/L_K^N$ value for the maximum disk model exceeds the value obtained from stellar models by more than 50%, i.e., the disk is apparently submaximal.

In the MOND model, the rotation curve of NGC 3992 is poorly described. Depending on the chosen parameter $a_0$, it satisfactorily approximates either the internal or external part of the observed rotation curve. The rotation curve shape at large distances from the center is consistent with the $a_0$ value adopted in MOND; however, the $M/L_K^M$ ratio for the disk as a whole exceeds the photometric estimate by more than 2.5 times, and the $M/L_K^N$ value for the maximum disk, by 70%.

**NGC 5055.** The Newtonian model well describes the long declining part of the rotation curve. In this case, same as for NGC 3521, the value $M/L_K^N = 0.29$ turns out to be lower than $M/L_K = 0.50$ derived from brightness and total color. The values of the parameter $a_0$ in the MOND models for this galaxy are consistent with the value adopted in MOND, and the agreement with the observed curve is no worse than in the Newtonian model. The MOND model also does not contradict the stellar population models: the $M/L^M$ value for the disk nearly coincides with the photometry-based value.

In order to compare the models for the seven galaxies under study, Table 4 lists the following: the photo-

**Table 3.** Parameters of Newtonian models and MOND

| Galaxy | $M/L_K$ | $M/L_K^N$ | $M/L_K^M$ | $a_0, 10^{-8}$ cm/s$^2$ |
|---|---|---|---|---|
| (1) | (2) | (3) | (4) | (5) |
| NGC 157 | 0.44 | 0.65 | 0.73 | $0.2 \pm 0.1$ |
| NGC 2841 | 0.76 | 0.96 | 1.46 | $1.3 \pm 0.2$ |
| NGC 2903 | 0.21–0.90 | 0.43–2.10 | 0.61–2.63 | $1.6 \pm 0.4$ |
| NGC 3031 | 0.80 | 0.80 | 0.88 | $0.6 \pm 0.2$ |
| NGC 3521 | 0.64 | 0.37 | 0.52 | $1.1_{+0.2}^{-0.4}$ |
| NGC 3992 | 0.64 | 1.07 | 1.82 | $1.2_{+0.2}^{-0.3}$ |
| NGC 5055 | 0.50 | 0.29 | 0.49 | $0.8 \pm 0.1$ |

The columns show: (1) the galaxy's name; (2) photometric estimate of the $M/L_K$ ratio for the stellar disk (see Section 4.1); (3) the disk's mass-to-luminosity ratio for the Newtonian maximum disk model; (4) the disk's mass-to-luminosity ratio for the MOND model; and (5) the MOND parameter estimate.

**Table 4.** Estimates of the galactic disk masses in different models

| Galaxy | $M_d$, $M_\odot$ | $M_d^N$, $M_\odot$ | $M_d^M$, $M_\odot$ |
|---|---|---|---|
| (1) | (2) | (3) | (4) |
| NGC 157 | $3.41 \times 10^{10}$ ($K$) | $5.02 \times 10^{10}$ | $5.64 \times 10^{10}$ |
| NGC 2841 | $9.97 \times 10^{10}$ (3.6 μm) | $1.28 \times 10^{11}$ | $2.24 \times 10^{11}$ |
| NGC 2903 | $1.37 \times 10^{10}$ (3.6 μm) | $3.68 \times 10^{10}$ | $4.41 \times 10^{10}$ |
| NGC 3031 | $7.61 \times 10^{10}$ (3.6 μm) | $7.61 \times 10^{10}$ | $8.37 \times 10^{10}$ |
| NGC 3521 | $1.03 \times 10^{11}$ (3.6 μm) | $5.61 \times 10^{10}$ | $8.21 \times 10^{10}$ |
| NGC 3992 | $5.10 \times 10^{10}$ (3.6 μm) | $8.51 \times 10^{10}$ | $1.45 \times 10^{11}$ |
| NGC 5055 | $1.11 \times 10^{11}$ (3.6 μm) | $4.81 \times 10^{10}$ | $9.04 \times 10^{10}$ |

The columns show: (1) the galaxy's name; (2) stellar disk mass according to the photometric data; (3) disk mass for the maximum disk model; (4) stellar disk mass in the MOND model.

metric estimate of the disk mass $M_d$ (the photometric band used to estimate the radial brightness profile is shown in brackets), the disk's maximum mass in the Newtonian model $M_d^N$, the disk mass in the MOND model $M_d^M$, and the corresponding parameter $a_0$ for which the observed decrease of rotation velocity agrees best with the model.

In the galaxies NGC 3521 and NGC 5055, the stellar disk mass taken from photometric data turned out to be greater than in the maximum disk's dynamic model. While the reason for this is not obvious, it may be asso- ciated, i.e., with an underestimation of the dust effect on the galaxy's color, or with the dynamic estimation inaccuracy of the disk mass due to the disk compo- nent's relatively low contribution to the rotation veloc- ity in the rotation curve maximum's region in these two galaxies. For other objects, the disk mass obtained from photometry is smaller than the mass determined from the rotation curve (by 40% on average), which is naturally expected for "submaximal" disks, since the maximum disk model gives only the upper limit of its mass.

Thus, the comparison of the Newtonian model and MOND for the seven galaxies under study showed that MOND describes the shape of the declining rotation curve no worse than Newtonian models in some cases; however, in three out of seven cases, the parameter $a_0$ value required for this differs from the one adopted in MOND. The difference is the greatest for NGC 157, where the rotation curve requires $a_0$ to be 6 times lower. The rotation curve's shape of this galaxy is not reproduced in the MOND variant under consider- ation.

In the MOND models at $a_0$ optimal values, the stellar disk mass for four of the seven galaxies (except NGC 5055, NGC 3031, and NGC 3521) turns out to be significantly higher, and for two galaxies (NGC 5055 and NGC 3521), on the contrary, lower than expected from the photometric estimates for the stellar population. In three cases (NGC 3521, NGC 3992, and NGC 5055), the stellar disk mass according to MOND exceeds the disk mass in the maximum disk model by more than 50%.

## 5. CONCLUSIONS

(1) The positions of 22 galaxies with declining rota- tion curves on the TF and BTF diagrams suggest that their rotation velocity corresponding to the curve maximum is normal for galaxies of similar luminosity or baryonic mass. At least for galaxies with the most significant decrease in the rotation curve at the periphery, it is apparently associated with a reduced rotation velocity of the disk's outer regions, and not with an abnormally high rotation velocity in the maximum region.

(2) There is no reason to believe that the rotation curve features of the galaxies under study noticeably affect their evolution. This is indirectly evidenced that the galaxies under consideration have no obvious mor- phological peculiarities, and their neutral hydrogen content does not differ from that of spiral galaxies with a similar angular momentum, which is considered proportional to the product $V_{\max} R_d$.

(3) The comparison of the Newtonian model with the MOND model for seven galaxies with the longest rotation curves concludes that the observed decrease in the rotation velocity at the periphery can be satis- factorily explained within the context of both models. However, in three cases out of seven, the required value of the parameter $a_0$ in MOND differs signifi- cantly from the one adopted in MOND (~$10^{-8}$ cm/s$^2$).

(4) In four of the seven galaxies under consider- ation (NGC 157, NGC 2841, NGC 2903, and NGC 3992), the disk mass in the MOND models is overes- timated in comparison with the values obtained from both disk photometry and the Newtonian maximum disk model. In order to agree with MOND, the $M/L$ ratio of the stellar disks in the $K$ band for these galaxies needs to significantly exceed the values derived from the stellar models. The most striking discrepancy with the MOND model is the galaxy NGC 157, whose rotation velocity decreases almost in half at the periphery. Neither the disk mass estimate, nor the rotation curve shape of this galaxy can be explained in MOND.

The MOND model does not directly contradict the observations in the analysis of statistical dependences (first of all, the dependence of the luminosity or baryonic mass of galaxies on the rotation velocity, or the relationship between the Newtonian and observed accelerations, see, e.g., [9]). However, the relationship between the stellar mass and the rotation velocity is also reproduced in numerical cosmological models of galaxy formation, i.e., in the context of Newtonian gravity [51]. It is important to note here that the MOND weak point is a prerequisite for the universality of the parameter $a_0$, which plays the role of a fundamental constant in this theory. Curiously, however, of the seven galaxies examined here, only NGC 157 has an estimate of $a_0$ very different than that expected in MOND.

Nevertheless, the examples presented in this study have shown that the baryonic mass estimates from MOND, in which there is no dark halo, poorly agree with the values of the total stellar mass obtained from the luminosity in the near-IR region. This demonstrates the evident difficulties faced by MOND when applied to specific objects.

## ACKNOWLEDGMENTS

The authors are grateful to the HyperLeda database [32].

*Translated by M. Chubarova*

SPELL: OK